\documentclass[10pt,twocolumn,journal]{IEEEtran}
\usepackage{amssymb}
\usepackage{amsfonts}
\usepackage{times,amsmath,epsfig}
\usepackage{color}
\usepackage{ulem}
\usepackage{latexsym}
\usepackage{multirow}
\begin{document}

\title{\huge Rate-Compatible LDPC Codes Based on Puncturing and Extension Techniques for Short Block Lengths}

\author{Jingjing Liu and Rodrigo C. de Lamare \vspace{-2em}
\thanks{Jingjing Liu and Rodrigo C. de Lamare \{jl622;rcdl500\}@york.ac.uk are with Department of Electronics, The University of York, Heslington, York, YO10 5DD, UK.}}% <-this % stops a s

\maketitle

\IEEEpeerreviewmaketitle

\begin{abstract}

In this paper, we investigate novel strategies for generating
rate-compatible (RC) irregular low-density parity-check (LDPC) codes
with short/moderate block lengths. We propose three puncturing and
two extension schemes, which are designed to determine the
puncturing positions that minimize the performance degradation and
the extension that maximize the performance. The first puncturing
scheme employs a counting cycle algorithm and a grouping strategy
for variable nodes having short cycles of equal length in the Tanner
Graph (TG). The second scheme relies on a metric called Extrinsic
Message Degree (EMD) and the third scheme is a simulation-based
exhaustive search to find the best puncturing pattern among several
random ones. In addition, we devise two layer-structured extension
schemes based on a counting cycle algorithm and an EMD metric which
are applied to design RC-LDPC codes. Simulation results show that
the proposed extension and puncturing techniques achieve greater
rate flexibility and good performance over the additive white
Gaussian noise (AWGN) channel, outperforming existing techniques.
\end{abstract}


\section{Introduction}
When the channel state information (CSI) is known at the transmit
end and the data transmission takes place over time-varying
channels, an error control coding scheme with a fixed code rate is
not regarded as the best solution. In such a situation, an
error-correction scheme with flexibility in code rates is desirable
since it is able to encode data at different rates depending on the
reliability of the channel. Higher rate codes are applied to achieve
higher data throughput if the channel condition is good, otherwise
lower rate codes are used to guarantee reliable transmission. Thus,
both capacity and reliability can be realized in such a scenario.
However, deploying many pairs of encoders and decoders is not
feasible in practical applications due to their high cost.
Rate-compatible (RC) codes refer to a family of codes where higher
rate codes are embedded in lower rate codes; in other words, the
factor graphs of higher rate codes are subgraphs of lower rate codes
\cite {Hagenauer}. For example, Lin and Yu \cite {Yu} designed an RC
coding scheme for a hybrid automatic repeat-request with forward
error correction (ARQ/FEC) system, where the transmitter keeps
sending additional redundant bits on request until the decoder
claims a successful decoding. Having been applied to convolutional
codes \cite{Hagenauer} and turbo codes \cite{Rowitch}, RC techniques
are proven not only to enhance system performance but also to
require only low hardware complexity thanks to the structure of a
single pair of encoder and decoder.

Low-Density Parity-Check (LDPC) codes were invented by Gallager
\cite {Gallager} and re-discovered by MacKay et al. \cite {Mackay}
as an advanced coding technique for Shannon capacity-approaching
performance over a variety of channels \cite{LDPCcapacity,Forney1}.
In \cite{Proietti,Ryne1}, finite-length LDPC codes were studied for
both ``waterfall" performance in the low signal-to-noise ratio (SNR)
region and error-floor behaviour in the high SNR region. Since
RC-LDPC codes were first considered in \cite {Li}, there has been a
fair amount of work in this area. Ha et al. \cite{Ha} derived
puncturing distributions via asymptotic analysis while assuming
infinite block length and no presence of short cycles. Later, in
\cite{Kim}, the authors focused on minimizing the number of
iterations required to recover punctured bits. Unlike \cite{Kim},
the work reported in \cite {Park} tries to maximize the minimum
reliability provided via check nodes. An efficiently-encodable
irregular LDPC codes along with a puncturing method were derived in
\cite{Mclaughlin}, where good performance can only be achieved via
puncturing degree-$2$ non-systematic bits. Also, in \cite{Bhushan},
protograph-based RC ensembles were implemented for hybrid ARQ
applications and systematic construction of punctured LDPC codes was
achieved via successive maximization, respectively. Other puncturing
methods can be found in \cite{Pishro-Nik,Vellambi} where the authors
proved the existence of a puncturing threshold with an improved
decoding algorithm, or enhanced the performance at high SNRs by
grouping nodes. On the other hand, extension methods, \cite{Li} and
\cite{Yazdani} have been applied by adding extra parity-check bits
to increase the size of the parity-check matrix of the mother code.
As a result, lower rate codes are generated based on a high rate
mother code.

We are interested in the design of RC-LDPC codes with reduced
performance degradation as compared to unpunctured codes with the
same rates. In our preliminary work \cite{Liu}, three puncturing
schemes {have been} proposed that are able to generate finite-length
RC-LDPC codes with good decoding performance at high rates (ranging
from $0.5$ to $0.9$). The first puncturing scheme is a
cycle-counting based (CC-based) technique that exploits the
algorithm reported in \cite{Halford} to determine the puncturing
pattern. Given a mother code and a target rate, variable nodes
having the largest number of girth-length cycles will be punctured
first, such that the decoding performance is expected to improve
while breaking the shortest cycles. Using a metric for evaluating
the extrinsic message degree (EMD), \cite{Tao} and
\cite{Vukobratovic}, a second scheme is called approximate cycle EMD
based (ACE-based) puncturing which selects the puncturing pattern by
considering the cycle length and graph connectivity simultaneously.
Additionally, a third scheme relies on a simulation-based greedy
search for the best puncturing pattern among many randomly generated
patterns. Based on the structure of short cycles and the ACE
spectrum, two extension schemes were  {devised in} \cite{Liu2}.
 {In this paper, we present puncturing and extension
techniques that are further enhanced and detailed of the methods
first reported in \cite{Liu,Liu2} along with technical analysis and
a comprehensive set of numerical results in terms of bit error rate
(BER), frame error rate (FER) and system throughput. The main
contributions can be summarized as:}

\begin{itemize}

\item  {We} present three puncturing schemes which depend upon
the counting cycle algorithm, ACE metric and an exhaustive search,
respectively. Further performance gain is observed regarding
 {the puncturing techniques reported in \cite{Liu}.}

\item We propose two extension schemes based on the structure of short cycles
and the ACE spectrum. The extension schemes proposed show an
improvement over the techniques described in \cite{Liu2}.

\item A comprehensive study of RC-LDPC codes is carried out
from the short cycles point of view. A set of puncturing/extension
strategies is made available to create RC-LDPC codes that are highly flexible in block
length, regularity (available for both regular and irregular codes)
and rate (across a wide range from $0.1$ to $0.9$).

\item A simulation study including extensive comparisons to
existing algorithms is presented. The performance of the
RC-LDPC codes proposed is shown with respect to BER, FER and throughput for a
type-\uppercase\expandafter{\romannumeral 2} hybrid automatic
repeat-request (ARQ) system.

\end{itemize}

The organization of this paper is as follows: Section
\ref{sec:SysMod} explains the  {system model and the basic
notation}. The proposed puncturing schemes and extension schemes are
detailed in Section \ref{sec:PropPunt} and Section \ref{sec:PropEx},
respectively. Section \ref{sec:Sim} presents simulation results with
explanations. Finally, Section \ref{sec:Sum} concludes this paper.

\section{System Model and Basic Notation}\label{sec:SysMod}

This section presents a system model for the proposed
puncturing and extension schemes, as well as the design strategy
behind them. All of the proposed techniques are based on cycle conditioning for each subgraph (puncturing) or extended graph
(extension). Note that traditional cycle conditioning has only
focused on eliminating short cycles in a Tanner Graph (TG) with very
long block sizes. But it has been proven that avoiding short cycles
alone is not enough to achieve good performance, particularly in the
error-floor region, \cite{Tao} and \cite{Vukobratovic}, and cycle
conditioning is a challenging task for finite-length LDPC codes.
Thus, in this work we derive a strategy to generate a family
of finite-length LDPC codes over the additive white Gaussian noise
(AWGN) channel.

\subsection{Construction of RC-LDPC Codes Using Puncturing}
%\begin{figure}[htb]
%\begin{minipage}[h]{1.0\linewidth}
%  \centering
%  \centerline{\epsfig{figure=puncmodel.eps,scale=0.7}} \vspace{-0em}\caption{System transmission model for puncturing.} \label{fig:puncmodel}
%\end{minipage}
%\end{figure}

Given a mother or parent LDPC code containing $K$ information bits,
the code rate is given by $R=K/N$ where $N$ is the block length.
Suppose that $\boldsymbol{m}$ represents the message from the
source, $\boldsymbol{c}$ denotes the encoded data, $\boldsymbol{c'}$
is the punctured data, and $\boldsymbol{\hat{m}}$ denotes the
estimate of the original message using a belief propagation (BP)
decoding algorithm given the unpunctured data $\boldsymbol{r}$.
Notice that the log-likelihood ratio (LLR) of a punctured bit is set
to $0$ at the beginning of the decoding process. Suppose that $P$
bits are punctured before transmission, so the resulting code rate
is given by $R' = K/(N-P)$ and the puncturing rate by $\rho=P/N$. We
assume that the decoder has perfect knowledge regarding the
puncturing pattern, {i.e.,} the position of punctured bits in a
codeword. Otherwise, some side information is needed to send the
puncturing pattern to the receiver end. Puncturing is a common and
simple method to construct RC codes, for which a higher rate is
achievable by means of removing a subset of encoded bits
$\boldsymbol{c}$ \cite {Pishro-Nik}. A randomly chosen puncturing
pattern \cite{Li} can be used to realize the rate compatibility at
the expense of severe performance degradation. Intentional
puncturing methods were investigated for short block LDPC codes in
\cite{Kim} - \cite{Mclaughlin} and \cite{Vellambi}, ranging from
asymptotic analysis to grouping and sorting variable nodes. In
contrast to those methods, the proposed puncturing schemes aim to
diminish the performance loss caused by puncturing from a cycle
distribution perspective,  {i.e.,} the puncturing pattern is
selected in the sense that the removed bits will break a certain
number of short cycles, which significantly improves the
connectivity of the TG.

\subsection{Construction of RC-LDPC Codes Using Extension}

\begin{figure}[htb]
\begin{minipage}[h]{1.0\columnwidth}
  \centering
  \centerline{\epsfig{figure=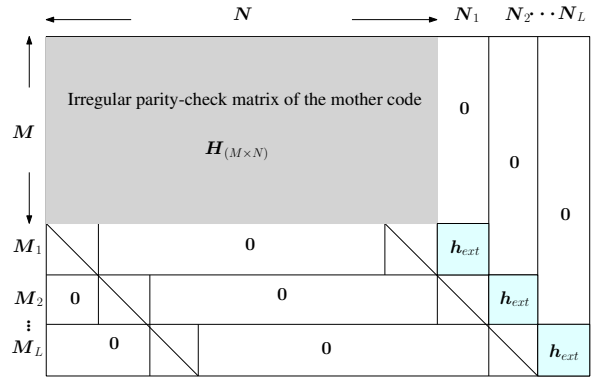,scale=0.4}} \vspace{-0em}\caption{To extend the parity-check matrix $\boldsymbol{H}_{(M \times N)}$ to $\boldsymbol{H}_{{\rm ext}(M_L \times N_L)}$ via a multi-level strategy.} \label{fig:Extending}
\end{minipage}
\end{figure}

The authors in \cite{Li} and \cite{Yazdani} state that extension is
another effective approach to constructing good RC-LDPC codes. We
also employ the idea of cycle conditioning to devise the proposed
extension schemes. The proposed extension framework is built as
shown in Fig. \ref {fig:Extending}, in which, starting from level
$1$ and running to level $L$, the current parity-check matrix is
extended in such a way that the same number of rows and columns are
added in each level. Consequently, the corresponding code rate
gradually reduces. Since $M_l=N_l=B (l=1, \ldots, L)$, the matrix
$\boldsymbol {h}_{\rm {ext}}$, along with two accompanying identity
matrices, is a $B \times B$ square matrix. Note that $\boldsymbol
{h}_{\rm {ext}}$ is fixed from one level to the next. In Fig. \ref
{fig:Extending}, the areas filled by ``$\boldsymbol{0}$" ensure the
sparseness of the extended parity-check matrix $\boldsymbol
{H}_{{\rm ext}}$, and the existence of identity matrices guarantees
a rather uniform degree of check node distribution as well as
creating sufficient dependency between the original matrix
$\boldsymbol {H}$ and the newly-extended matrix $\boldsymbol
{H}_{{\rm ext}_l}$. Our framework is very similar to that in \cite
{Yazdani}, which enables fast linear-time encoding as
 {the} matrix $\boldsymbol {H}_{{\rm ext}}$ is always
obtained in systematic form by using Gauss-Jordan elimination.
Furthermore, the proposed extension schemes have two extra features:
1) possible cycles of length $4$ are avoided by not putting two
identity matrices together; 2) more importantly, the submatrices
$\boldsymbol {h}_{{\rm ext}}$ are carefully chosen with cycle
conditioning for each subgraph. If $\boldsymbol {I}$ denotes the
identity matrix and $\boldsymbol {G}$ denotes the generator matrix
of the mother code, $\boldsymbol {G}_1, \ldots, \boldsymbol {G}_L$
are systematically transformed from the extended counterparts of
$\boldsymbol {H}_{{\rm ext}1}, \ldots, \boldsymbol {H}_{{\rm
ext}L}$. Similar to the puncturing model, some side information is
needed at the receiver end to indicate the desired rate and
corresponding parity-check matrix.

%\begin{figure}[htb]
%\begin{minipage}[h]{1.0\linewidth}
%  \centering
%  \centerline{\epsfig{figure=systemmodel.eps,scale=0.7}} \vspace{-0em}\caption{System transmission model for proposed extension schemes.} \label{fig:sysmodel}
%\end{minipage}
%\end{figure}

\section{Proposed Puncturing Techniques}\label{sec:PropPunt}

Inspired by the cycle-conditioning and ACE metric, in this section
we introduce the proposed puncturing schemes,  {i.e.,} CC-based
puncturing, ACE-based puncturing, and simulation-based puncturing.
The first two methods are developed using the counting cycle
algorithm and the ACE metric, while the last one is based on an
exhaustive search. As mentioned in the last section, all the
proposed puncturing techniques can be applied offline (independent
of the data transmitted), and without any side information the
puncturing patterns need to be stored at both the transmitter and
receiver ends. Unlike the preliminary results reported in
\cite{Liu}, we have replaced the cycle counting algorithm
\cite{Halford} with a more efficient algorithm \cite{Karimi},
modified the puncturing order of the proposed ACE-based scheme, and
employed  {an} improved PEG code (ACE PEG) \cite{Xiao} as the mother
code.

\subsection{CC-Based Puncturing Scheme}

The proposed counting cycle (CC)-based puncturing technique is
developed based on the counting cycle algorithms \cite {Halford} and
\cite {Karimi}. The former algorithm employs matrix multiplications
while the latter takes advantage of the message-passing nature of BP
decoding. Given the same TG, we have verified that both algorithms
produce similar results for counting cycles of length $g$ and $g+2$,
where $g$ is the girth. But the algorithm in \cite{Karimi} has much
lower complexity ($\mathcal{O}(g|E|^2)$) than its counterpart \cite
{Halford} ($\mathcal{O}(gN^3)$), especially for graphs with large
sizes. Provided with the cycle distribution, the objective is to
select an ideal puncturing pattern that can break as many
girth-length cycles as possible, which may reduce the performance
degradation  {introduced} by puncturing. The idea behind the
proposed algorithm is inspired by the fact that the existence of
short cycles creates a statistical dependency between the extrinsic
messages being exchanged in the current decoding iteration, such
that the extrinsic messages for the next iteration will,
inaccurately, have high reliability.

According to the PEG algorithm \cite{Hu}, high degree nodes are
placed in the leftmost positions of $\boldsymbol H_{(M\times N)}$
that correspond to information bits, as they provide more protection
for the original data. Following this design rule we only puncture
the set of variable nodes $s_j\in {V_s}$  {, where $K+1 \leq j \leq
N$}. Define the vector $\boldsymbol
c_{s_{j}}=\{N_g,N_{g_2},N_{g_4}\}^T$ whose element refers to the
number of $g-{\rm cycles}$, $(g+2)-{\rm cycles}$ and $(g+4)-{\rm
cycles}$ passing through a variable node $s_j$. Any $s_j (K+1 \leq j
\leq N)$ will be included as a punctured candidate if $N_g\neq 0$.
For each candidate node, another vector $\boldsymbol {v}_{g\_{s_j}}$
is formed as:
\begin{equation} \label{3}
\centering {\boldsymbol
{v}_{g\_{s_j}}=\{v_{g\_{s_0}},v_{g\_{s_1}},\ldots,
v_{g\_{s_{N-1}}}\}^T,~~~ s_j\in V_c},
\end{equation}
where  {the} entries represent the number of cycles of length $g$
that $s_j$ has, and are arranged in decreasing order. Similarly, we
can also define $\boldsymbol {v}_{{g_2}{\_{s_j}}}$ or $\boldsymbol
{v}_{{g_4}{\_{s_j}}}$ if necessary. There are two criteria to
determine the set of punctured nodes: 1) to find variable nodes
having the shortest cycles passing through; 2) to find variable
nodes having more such cycles than others. In addition, we also
tried to arrange the entries of (\ref{3})  {in a reverse manner,
i.e., we start by puncturing} the variable nodes having the least
number of cycles of length $g$. But with such a formation, the
performance deteriorates dramatically. If  {the} candidates on the
$g-{\rm cycles}$ are less than $P$, we puncture $P$  nodes at first
then arrange the rest of the candidates with respect to the $(g+2)-$
and $(g+4)-{\rm cycles}$.  {This} situation rarely occurs in
practice unless an unreasonable puncturing rate $\rho$ is given.
Compared to random puncturing schemes, CC-based puncturing requires
more computational complexity due to the cycle counting algorithm.
On the other hand, CC-based puncturing has been verified
 {to} significantly outperforming random puncturing
techniques \cite{Liu}. Obviously, the complexity of CC-based
puncturing is mainly increased by counting short cycles. It is worth
noting that the practical complexity is lower than
$\mathcal{O}(g|E|^2)$ since, most of the time only a counting cycle
of length $g$ is required. The proposed counting cycle (CC)-based
puncturing technique can be summarized  {as}:

\textbf {Step 1:} given  {the} block size $N$,  {the} rate $R$ and
the degree distribution, generate the parity-check matrix of
 {the}
mother code $\boldsymbol H_{(M\times N)}$ by using  {the improved PEG algorithm} \cite{Xiao};\\

\textbf {Step 2:} for $\boldsymbol H_{(M\times N)}$ compute $g-{\rm
cycle}$, $(g+2)-{\rm cycle}$ and $(g+4)-{\rm cycle}$ with
respect to variable nodes $s_j\in {V_s}$ where $(K+1 \leq j \leq N)$;\\

\textbf {Step 3:} based on the knowledge from Step 2, define
vector $\boldsymbol c_{s_{j}}=\{N_g,N_{g_2},N_{g_4}\}^T$ for every
variable node $s_j$.
If $N_g\neq 0$, $s_j$ is picked as one of the punctured candidates;\\

\textbf {Step 4:} for all the candidates chosen in Step 3, define
the vector $\boldsymbol{v}_{g\_{s_j}}$ $(s_j\in V_s)$. Puncture the
first $P$ candidates in $\boldsymbol{v}_{g\_{s_j}}$.\\

\begin{table*}
\centering
    \caption{Cycle Distributions of code A before and after CC-based
    puncturing}
    \label{tab:cycledisa}
\begin{tabular}{|c|c|c|c|c|c|c|c|}
\hline \multirow{2}{*}{} & \multicolumn{3}{|c|}{Mother code
$N=1000$} & \multirow{2}{*}{}&
\multicolumn{3}{|c|}{Punctured code $N=800$} \\
\cline{2-4} \cline{6-8}
& $N_c$ & $\mu_c$ &  $\sigma_c$ & & $N_c$ & $\mu_c$ & $\sigma_c$ \\
\hline
$c=8$ & 513 & 3.5  & 5.45 & $c=8$ & 295 & 28.1 & 32.17\\
\hline
$c=10$ & 18553 & 148.52 & 143.31 & $c=10$ & 12853 & 385.53 & 372.31\\
\hline
$c=12$ & 198607 & 2482.6 & 2298.3 & $c=12$ & 191287 & 4684 & 4558.3\\
\hline
\end{tabular}
\end{table*}

Now we illustrate how CC-based puncturing affects the cycle
distribution as well as  {the} overall performance. As for a cycle
of length $c$, the cycle distribution is defined as $(N_c, \mu_c,
\sigma_c)$ where $N_c$ denotes the number of cycles of length $c$,
while $\mu_c$ and $\sigma_c$ denote the mean and standard deviation
of $c$-length cycles with respect to the variable nodes. By way of
example, we use an irregular TG of code A in Section
\uppercase\expandafter{\romannumeral 5}. Table \ref {tab:cycledisa}
shows the cycle distributions of the mother code and the punctured
code. Applying CC-based puncturing, short cycles of length $g=8$ are
reduced by $42\%$ while short cycles of length $10$ and $12$ are
reduced by $30\%$ and $3\%$, respectively. Even if the number of
girth-length cycles diminishes, it is worth noticing that the cycle
distribution becomes less uniform after puncturing. In
\cite{Halford}, the authors suggest that with the same girth
 {codes with a rather uniform cycle distribution
perform better than codes with} a non-uniform cycle distribution. As
a consequence, the proposed CC-based puncturing removes a fair
number of cycles of girth length but also damages the inherent
connectivity of the TG. Based on this fact, we are motivated to
devise more advanced puncturing scheme as a sequel.

\subsection{ACE-Based Puncturing Scheme}

The second puncturing algorithm proposed is an improved
version of the CC-based puncturing scheme, which is called an ACE-based
puncturing algorithm, thanks to the employment of an ACE metric. ACE-based puncturing strives to remove a certain of short
cycles and simultaneously maintain good graph connectivity.
Since not all short cycles of the same length are equally
detrimental to iterative decoding, the ACE metric \cite{Tao}
and ACE spectrum \cite{Vukobratovic} were introduced to evaluate
the consequences of short cycles with a certain length in a TG. For a cycle
$\mathfrak{C}$ and a corresponding set of variable nodes $ V_{\mathfrak{C}}$,
all the edges connected to $\mathfrak{C}$ can be categorized into
three groups \cite{Vukobratovic}, in which $ E_{\rm ext}( V_{\mathfrak{C}})$, including extrinsic edges incident
to those check nodes with a single connection to $
V_{\mathfrak{C}}$, is expected to be large so that short cycles will possess more singly
connected extrinsic edges, which decreases the probability of
cycles forming a small stopping \cite{Proietti} or trapping
set \cite {Tom}. For short cycles of the same length, a larger
ACE value indicates better connections to the rest of the graph.
Here we define the average ACE value regarding a variable node $s_j$
contained in $N_g$ cycles of length $g$ as:
\begin{equation} \label{av}
\centering {\alpha_{g}=1/{N_g}\sum_{1}^{N_g} \epsilon_{ACE}},
\end{equation}
where $\alpha_{g2}$ and $\alpha_{g4}$ are defined with respect to
cycles of length $g+2$ and $g+4$. Moreover, for each $s_j \in {V_s}$
where $(K+1 \leq j \leq N)$, $\alpha_{s_j}$ is defined as:
\begin{equation} \label{asj}
\centering {\alpha_{s_j}=\min\{\alpha_{g}, \alpha_{g2},
\alpha_{g4}\}}.
\end{equation}
Compared to the work reported in \cite{Liu}, the ACE puncturing proposed has the following three improvements: 1) the puncturing
ordering is adjusted to consider the connectivity of cycles prior to
their length; 2) a new code design \cite{Xiao} for generating the mother code makes
indexing ACE values more conveniently; 3) the combination of a new
design method and ordering leads to improved performance for both mother code
and punctured code. The proposed ACE puncturing can be depicted as
follows:

\textbf {Step 1:} given  {the} block size $N$,  {the} rate $R$, and
the degree distributions, generate the parity-check matrix for the
mother code
$\boldsymbol H_{(M\times N)}$ by using the improved PEG \cite{Xiao};\\

\textbf {Step 2:} for $\boldsymbol H_{(M\times N)}$ compute $g-{\rm
cycle}$, $(g+2)-{\rm cycle}$ and $(g+4)-{\rm cycle}$ for the variable nodes $s_j\in {V_s}$ where $(K+1 \leq j \leq N)$;\\

\textbf {Step 3:} with the knowledge from Step 2, define the
vector $\boldsymbol c_{s_{j}}=\{N_g,N_{g_2},N_{g_4}\}^T$ for every
variable node $s_j$ ($s_j\in { V_s}$). Calculate $\alpha_{s_j}$ using (\ref{av}) and (\ref{asj});\\

\textbf {Step 4:} find the set of puncturing candidates
${\boldsymbol {w}=\{\alpha_{s_0}, \alpha_{s_1},\ldots,
\alpha_{s_{N-1}}\}}^T$ by sorting $\alpha_{s_j}$ in increasing
order;\\

\textbf {Step 5:} puncture the
first $P$ candidates in $\boldsymbol {w}$.\\

\begin{table*}
\centering
    \caption{Cycle Distributions of code A before and after ACE-based
    puncturing}
    \label{tab:cycledisb}
\begin{tabular}{|c|c|c|c|c|c|c|c|}
\hline \multirow{2}{*}{} & \multicolumn{3}{|c|}{Mother code
$N=1000$} & \multirow{2}{*}{}&
\multicolumn{3}{|c|}{Punctured code $N=800$} \\
\cline{2-4} \cline{6-8}
& $N_c$ & $\mu_c$ &  $\sigma_c$ & & $N_c$ & $\mu_c$ & $\sigma_c$ \\
\hline
$c=8$ & 513 & 3.5  & 5.45 & $c=8$ & 402 & 10.88 & 15.7\\
\hline
$c=10$ & 18553 & 148.52 & 143.31 & $c=10$ & 15236 & 175.4 & 200.37\\
\hline
$c=12$ & 198607 & 2482.6 & 2298.3 & $c=12$ & 172547 & 2787.25 & 2523.44\\
\hline
\end{tabular}
\end{table*}

We use Table \ref {tab:cycledisb} to illustrate the change in the
cycle distribution after running the ACE puncturing scheme. Compared
to the results from Table \ref {tab:cycledisa}, ACE puncturing is
able to maintain a relatively uniform cycle distribution by first
removing the variable nodes which get involved with longer cycles
but have low ACE values. From the decoding point of view, in a
subgraph with good connectivity, the LLR of punctured bits is
expected to be recovered within a few iterations, even though there
might be other punctured bits in the same neighborhood. On the other
hand, unlike CC puncturing, ACE puncturing does not work for regular
codes since all the ACE values of variable nodes are identical. In
that case, it is impossible to consider puncturing priority with the
ACE metric.

\subsection{Simulation-Based Puncturing Scheme}

The last puncturing scheme proposed is developed on the basis of an
exhaustive search among a large number of random puncturing patterns. Then, the
best puncturing pattern is determined simply by choosing the one
having the best average BER performance. At the receiver end, in
order to find the best pattern, we need to send a training sequence then compute the average BER values at
$T$ SNR points for each puncturing pattern. For $Q$ possible patterns the
best pattern $\boldsymbol {p}_{\rm opt}$ is selected as:

\begin{equation} \label{12}
\centering \boldsymbol {p}_{\rm
opt}=\arg\min_{q}\frac{1}{rT}\sum_{i=1}^{r}\sum_{t=1}^{T}{\rm
BER}(\boldsymbol {p}_q),~~~ q=1, \ldots, Q.
\end{equation}

The proposed simulation-based (SIM-based) algorithm can be described as follows:

\textbf {Step 1:} given  {the}block size $N$,  {the} rate $R$ and
the degree distributions, generate the parity-check matrix of the
mother code
$\boldsymbol H_{(M\times N)}$ by using the improved PEG \cite{Xiao};\\

\textbf {Step 2:} for the desired rate $R'$, randomly generate $Q$
puncturing patterns represented by a row vector $\boldsymbol
{p}_q$ where $q=1, \ldots, Q$, in each of which $P$ bits are randomly punctured from the encoded data;\\

\textbf {Step 3:} for each
pattern in $\boldsymbol {p}_q$, send a training sequence of length $1,000$ at $T$ SNR points then calculate BER values;\\

\textbf {Step 4:} after running $r$ repetitions, for all $Q$ patterns calculate an average based on accumulated BER values;\\

\textbf {Step 5:} select the best puncturing pattern $\boldsymbol
{p}_{\rm opt}$ among
$\boldsymbol {p}_1, \ldots,\boldsymbol {p}_Q$ by choosing the pattern with the minimum average BER.\\

From \eqref{12}, it is apparent that given a desired rate $R'$ it is possible to obtain the optimal pattern
$p_{\rm opt}$ when all $\frac {N!}{M!(N-M)!}$ possible
puncturing patterns are considered, which seems infeasible in
practice. Since the quality of the best pattern
$\boldsymbol {p}_{\rm opt}$ depends on $Q$, the last proposed
puncturing scheme offers flexible trade-offs between performance
and the number of candidate patterns. In \cite{Liu}, SIM-based puncturing always outperforms CC-based
puncturing and ACE-based puncturing. Nevertheless, with the additional improvement,
ACE-based puncturing is able to provide at least comparable performance to
SIM-based puncturing, even when we increase $Q$ to $500$.

\section{Proposed Extension Techniques}\label{sec:PropEx}

In this section, we investigate another pathway to generate RC LDPC codes, i.e. extension techniques, and two proposed
schemes are explained in the sequel. An extension framework introduced in \cite {Yazdani} is
exploited which enables fast encoding and off-line operation for the proposed extension schemes. To refine the
techniques described in \cite{Liu2}, we replace the cycle
counting algorithm \cite{Halford} by a more efficient algorithm \cite{Karimi}, further develop the design process, and utilize the improved PEG code (ACE PEG) \cite{Xiao} as the mother code.

\subsection{Counting-cycle based extension}

The first extension scheme proposed is the counting cycle (CC)-based
extension which employs an algorithm for counting short cycles in
order to select  {an} extension submatrix $\boldsymbol {h_{ext}}$
among $S$ candidates. We set the parameter $S$ to equal the number
of desired extension rates, e.g. $S=3$ if $R=5/10$, $R_1=5/12$ and
$R_2=5/13$, where $R$ is the rate of the mother code. In this case,
three distinct submatrix candidates are constructed using the ACE
PEG algorithm \cite{Xiao} with different degree distributions, which
are derived via density evolution (DE) \cite{LDPCcapacity} and
provided with the maximum variable nodes' degree $d_{v_{\rm max}}$
and check nodes' degree $d_{c_{\rm max}}$. As per the extended part
in Fig. \ref{fig:Extending}, the $B \times B$ submatrix $\boldsymbol
{h}_{\rm {ext}}$ is very likely to have many short cycles while the
rest of  {the} extended part can be proved cycle free. Define
$g_{h(s)}(s=1, 2, \ldots, S)$ as the local girths for each candidate
submatrix, and $N_{g(s)}$ as the number of cycles of length
$g_{h(s)}$ corresponding to each $\boldsymbol{h}_s (s=1, 2, \ldots,
S)$. After running the counting cycle algorithm \cite{Karimi}, we
select the candidate submatrix with the largest $g_{h}$ and the
smallest $N_{g}$ as $\boldsymbol {h}_{ext}$. As a result, the
CC-based extension scheme maximizes the local girth $g_h$ of
$\boldsymbol {h}_{ext}$, and the selected $\boldsymbol {h}_{ext}$
has the smallest number of length-$g_h$ cycles. The algorithm flow
of the proposed CC-based extension is summarized as follows:

\textbf {Step 1:}  {given} the parity-check matrix $\boldsymbol
H_{(M \times N)}$ and the desired code rates $R_1, R_2, ..., R_L$,
determine the number of extension levels $L$ which ensures
$M_l=N_l=B (l=1, \ldots, L)$;\\

\textbf {Step 2:} set $S=L+1$, given $d_{v_{\rm max}}$ and
$d_{c_{\rm max}}$ derive $S$ degree distributions according to DE;\\

\textbf {Step 3:} based on Step 2, construct $S$ candidates for $B
\times B$ submatrices by using the improved PEG algorithm \cite{Xiao};\\

\textbf {Step 4:} for each submatrix candidate compute the $g_{h(s)}$ and $N_{g(s)}$ of each subgraph;\\

\textbf {Step 5:} choose the candidate with the largest
$g_{h(s)}$ and the smallest $N_{g(s)}$ as $\boldsymbol {h}_{\rm {ext}}$;\\

\textbf {Step 6:} for $1 \leq l \leq l$ gradually extend
$\boldsymbol H_{(M \times N)}$ to $\boldsymbol {H}_{{\rm {ext}}(L)}$
by adding zero entries, identity matrices, and $\boldsymbol {h}_{\rm
{ext}}$ as in Fig. \ref
{fig:Extending}.\\

\subsection{ACE-based extension}

The second scheme proposed is an ACE-based extension scheme. Unlike CC-based extension,
the candidate submatrix with the largest $\alpha(g_h)$ will be selected as the
$\boldsymbol {h}_{\rm {ext}}$, where $\alpha(g_{h(s)})$ is the
average ACE spectrum with respect to $N_{g(s)} (s=1, 2, \ldots, S)$,
as defined in (\ref{av}). Similar to ACE-based puncturing, it is straightforward to compute $\alpha(g_h)$ if the submatrix candidates are created using the ACE PEG
algorithm, \cite{Xiao}. The proposed ACE-based extension is described in the following:

\textbf {Step 1:}  {given} the parity-check matrix $\boldsymbol
H_{(M \times N)}$ and the desired code rates $R_1, R_2, ..., R_L$,
determine the number of extension levels $L$ which ensures
$M_l=N_l=B (l=1, \ldots, L)$;\\

\textbf {Step 2:} set $S=L+1$, given $d_{v_{\rm max}}$ and
$d_{c_{\rm max}}$ derive $S$ degree distributions according to DE;\\

\textbf {Step 3:} based on Step 2, construct $S$ candidates for $B
\times B$ submatrices by using the improved PEG algorithm \cite{Xiao};\\

\textbf {Step 4:} for each submatrix candidate compute $g_{h(s)}$ and $N_{g(s)}$ for each subgraph;\\

\textbf {Step 5:} provided with $g_{h(s)}$ and $N_{g(s)}$ in Step 3,
calculate $\alpha(g_{h(s)})$;\\

\textbf {Step 5:} choose the candidate with the largest
$\alpha(g_{h(s)})$ as $\boldsymbol {h}_{\rm {ext}}$;\\

\textbf {Step 6:} for $1 \leq l \leq l$ gradually extend
$\boldsymbol H_{(M \times N)}$ to $\boldsymbol {H}_{{\rm {ext}}(L)}$
by adding zero entries, identity matrices, and $\boldsymbol {h}_{\rm
{ext}}$ as in Fig. \ref
{fig:Extending}.\\

\section{Simulation Results}\label{sec:Sim}

First, this section presents numerical results corresponding to the
three forms of proposed puncturing and two proposed extension
algorithms, respectively. Then, joint comparisons are carried out of
the puncturing and extension schemes at different rates, which shows
that the former performs better at higher rates while the latter is
superior at lower rates. In all the simulations, mother codes are
finite-length irregular LDPC codes generated by the improved PEG
algorithm \cite{Xiao}. Code A has blocklength of $N=1,000$, code
rate $R=0.5$ and degree distributions $\mu(x)=0.21\times
x^5+0.25\times x^3+0.25\times x^2+0.29\times x$, $\nu(x)=x^5$. Code
B has blocklength $N=2,000$, code rate $R=0.4$ and degree
distributions $\mu(x)=0.45\times x^9+0.26\times x^2+0.29\times x$,
$\nu(x)=x^5$. The decoder applies the standard BP algorithm in the
logarithm domain.

\begin{figure}[!htb]
\begin{center}
\def\epsfsize#1#2{0.95\columnwidth}
\epsfbox{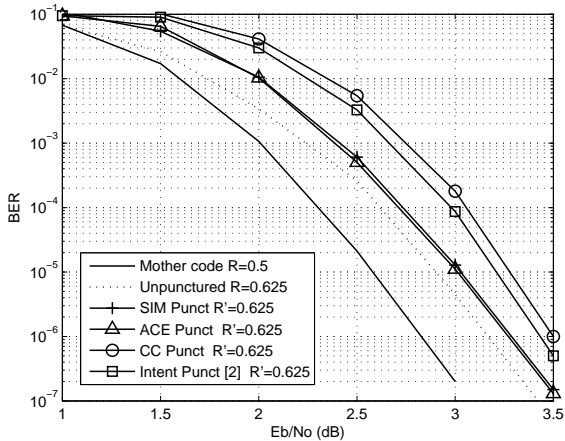} \vspace{-0.5em}\caption{Comparisons of the
proposed puncturing schemes with an existing puncturing scheme
\cite{Vellambi} with respect to BER performance, where code A is the
mother code.}\label{fig:irregularber}
\end{center}
\end{figure}

%\begin{figure}[!htb]
%\begin{center}
%\def\epsfsize#1#2{1.0\columnwidth}
%\epsfbox{ACEFER.eps}
%\vspace{-0.5em}\caption{Comparison of the proposed ACE
%puncturing scheme with an existing puncturing scheme in \cite{Vellambi}
%with respect to FER performance, where code A is the mother
%code.}\label{fig:irregularfer}
%\end{center}
%\end{figure}

\begin{figure}[!htb]
\begin{center}
\def\epsfsize#1#2{0.95\columnwidth}
\epsfbox{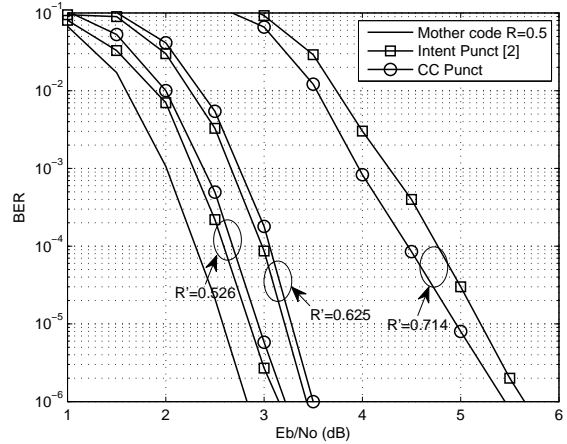} \vspace{-0.5em}\caption{Comparison of the
proposed CC-based puncturing with an existing puncturing scheme
\cite {Vellambi} at different resulting rates. $R$ is the rate of
the mother code, A, and $R'$ is the resulting
rate.}\label{fig:inrates}
\end{center}
\end{figure}

To test the proposed puncturing schemes, we first choose code A as
the mother code, then compare the performance to that of the
puncturing technique reported in \cite{Vellambi}. In this scenario,
the decoder runs a maximum of $60$ decoding iterations. Fig.
\ref{fig:irregularber} shows a BER performance comparison of the
three proposed schemes with the existing method \cite{Vellambi}, in
which the resulting rate $R'$ is $0.625$ and the puncturing rate
$\rho=0.2$, such that $200$ bits are punctured prior to
transmission. It is clear to see that the proposed ACE-based
puncturing significantly outperforms the existing method as well as
CC-based puncturing and slightly surpasses SIM-based puncturing. For
comparison purposes, we also include unpunctured irregular LDPC
code, with $N=800$, $R=0.625$, which has the same degree
distributions as the mother code A. Notice that the performance gap
between ACE-based puncturing and the unpunctured code is less than
$0.2$ dB at BER of $10^{-6}$. Additionally, more results over a
range of puncturing rates are shown in Figs. \ref{fig:inrates}, in
which CC-based puncturing begins to outperform \cite{Vellambi} by
$0.25$ dB at BER $10^{-5}$ after the resulting rate of $0.625$.
Since the performance of \cite{Vellambi} dramatically degrades
beyond the puncturing threshold $R'=0.65$, an additional algorithm
needs to be applied to achieve good performance at higher rates.

%\begin{figure}[!htb]
%\begin{center}
%\def\epsfsize#1#2{1.0\columnwidth}
%\epsfbox{revisedacerates.eps} \vspace{-0.5em}  \caption{Comparisons
%of the proposed ACE-based puncturing with an existing puncturing scheme
%\cite {Vellambi} at different resulting rates. $R$ is the rate of
%the mother code, A, and $R'$ is the resulting
%rate.}\label{fig:acerates}
%\end{center}
%\end{figure}

%\begin{figure}[!htb]
%\begin{center}
%\def\epsfsize#1#2{1.0\columnwidth}
%\epsfbox{revisedSMrates.eps} \vspace{-0.5em}  \caption{Comparison
%of the proposed SIM-based puncturing with an existing puncturing scheme
%\cite {Vellambi} at different resulting rates. $R$ is the rate of
%the mother code, A, and $R'$ is the resulting
%rate.}\label{fig:SMrates}
%\end{center}
%\end{figure}

\begin{figure}[!htb]
\begin{center}
\def\epsfsize#1#2{0.95\columnwidth}
\epsfbox{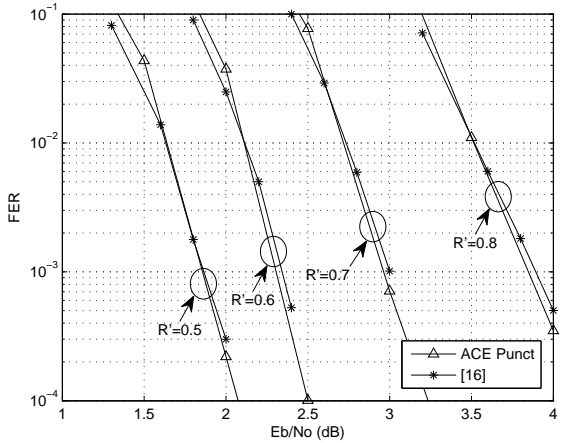} \vspace{-0.5em}  \caption{The comparison of the
puncturing FER performance between ACE-based puncturing and
\cite{Mclaughlin}. The puncturing rates $R'$ are $0.5$, $0.6$, $0.7$
and $0.8$. We use the mother code B with block length $N=2,000$ and
rate $R=0.4$.}\label{fig:codebfer}
\end{center}
\end{figure}

To test the performance of  {the proposed puncturing techniques} in
the type-\uppercase\expandafter{\romannumeral 2} ARQ system, in Fig.
\ref{fig:codebfer} we compare the proposed ACE-based puncturing with
the puncturing scheme in \cite{Mclaughlin} in terms of FER
performance. In this case, code B is used as the mother code and the
maximum number of decoding iterations is increased to $200$. From
 {Fig. \ref{fig:codebfer}}, we see that the puncturing
scheme of \cite{Mclaughlin} works better in the low SNR region but
is outperformed by ACE-based puncturing in the high SNR region. The
advantages of ACE-based puncturing over \cite {Mclaughlin} are as
follows:
\begin{itemize}
\item the method \cite{Mclaughlin} is easy to implement  {in} hardware, thanks to a specific code
structure. But it usually has to compromise on the optimal degree
distribution so as to fulfil the design requirement that may affect
the performance. ACE-based puncturing is a more general technique,
and can be applied to any irregular mother
code;\\
\item the ACE-based method aims to cover a good ACE spectrum via puncturing so graph connectivity is always taken into account at each rate.
Due to the design nature of \cite{Mclaughlin}, one has to maximize the
number of degree $2$ variable nodes whose ACE value is $0$. Once a
cycle is formed, that will severely reduce
performance, especially in the high SNR region;\\
\item the best puncturing performance for \cite{Mclaughlin} results from
$N_v(2)=M-1$ where $N_v(2)$ is the number of degree $2$ nodes.
However, this requirement is difficult to realize for a
mother code with a low rate; \\
\item the ACE scheme is expected to achieve any puncturing rate
without limitations, while method \cite{Mclaughlin} always has a puncturing
threshold of $R_H=K/(N-N_v(2))$, above which one can only use random
puncturing to achieve a higher rate.
\end{itemize}

%\begin{figure}[!htb]
%\begin{center}
%\def\epsfsize#1#2{1.0\columnwidth}
%\epsfbox{codebber.eps} \vspace{-0.5em}\caption{A comparison of the
%puncturing BER performance between ACE-based
%puncturing and \cite{Mclaughlin}. The puncturing rates $R'$ are $0.5$, $0.6$,
%$0.7$ and $0.8$. Here we use code B as the mother
%code.}\label{fig:codebber}
%\end{center}
%\end{figure}

As for the extension schemes, we compare the proposed extension techniques
with the technique reported in \cite {Yazdani}. In the following
simulations, we use the mother code, C, constructed by improved PEG \cite{Xiao} with block length $N=1,000$, $R=5/10$ and degree distributions
$\mu(x)=0.438x^6+0.416x^2+0.315x$ and $\nu(x)=0.561x^6+0.438x^5$.
The decoder terminates after a maximum of $100$ iterations.

\begin{figure}[!htb]
\begin{center}
\def\epsfsize#1#2{0.95\columnwidth}
\epsfbox{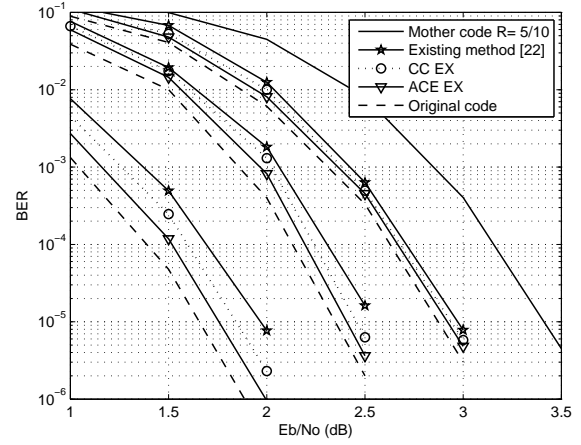} \vspace{-0.5em}\caption{Comparison of the
proposed extension schemes with another existing scheme
\cite{Yazdani} at different rates for irregular PEG LDPC codes. The
mother code corresponds to the rightmost curve with $N_0=1,000$ and
$R=5/10$. For other codes the rates from left to right are $5/14$,
$5/13$, $5/12$.}\label{fig:compare}
\end{center}
\end{figure}

%\begin{figure}[!htb]
%\begin{center}
%\def\epsfsize#1#2{1.0\columnwidth}
%\epsfbox{codelength1.eps}
%\vspace{-0.5em}\caption{Comparisons of the proposed
%extension schemes with another existing scheme \cite{Yazdani} for
%different block lengths at rate $5/13$ and SNR= $2$ dB for irregular
%LDPC codes. }\label{fig:codelength}
%\end{center}
%\end{figure}

In Fig. \ref {fig:compare}, we compare the proposed CC-based and ACE-based
extension algorithms with the existing extension method \cite
{Yazdani}. From rate $5/10$ to $5/14$, the extension operates at
three levels and $100$ bits are added per level. Notice that all the
degree distributions are constrained by $d_{v_{\rm max}} \leq 7$. In
Fig. \ref {fig:compare}, both proposed schemes outperform the
existing method at different rates, and the performance gap increases as
more parity bits are inserted.

\begin{figure}[!htb]
\begin{center}
\def\epsfsize#1#2{0.95\columnwidth}
\epsfbox{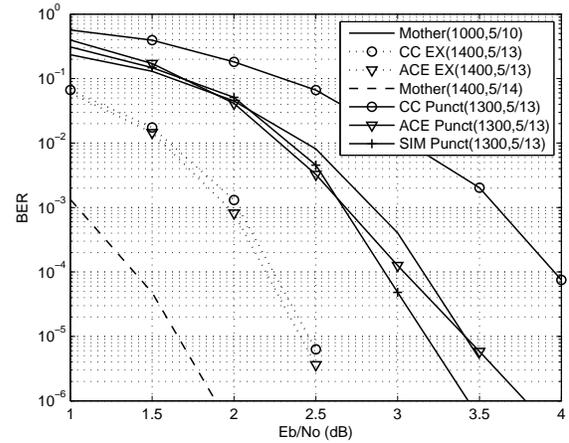} \vspace{-0.5em}\caption{Comparison of the
proposed extension schemes with the proposed puncturing schemes at a
low rate, $5/13$, for irregular LDPC code. }\label{fig:RClow}
\end{center}
\end{figure}

\begin{figure}[!htb]
\begin{center}
\def\epsfsize#1#2{0.95\columnwidth}
\epsfbox{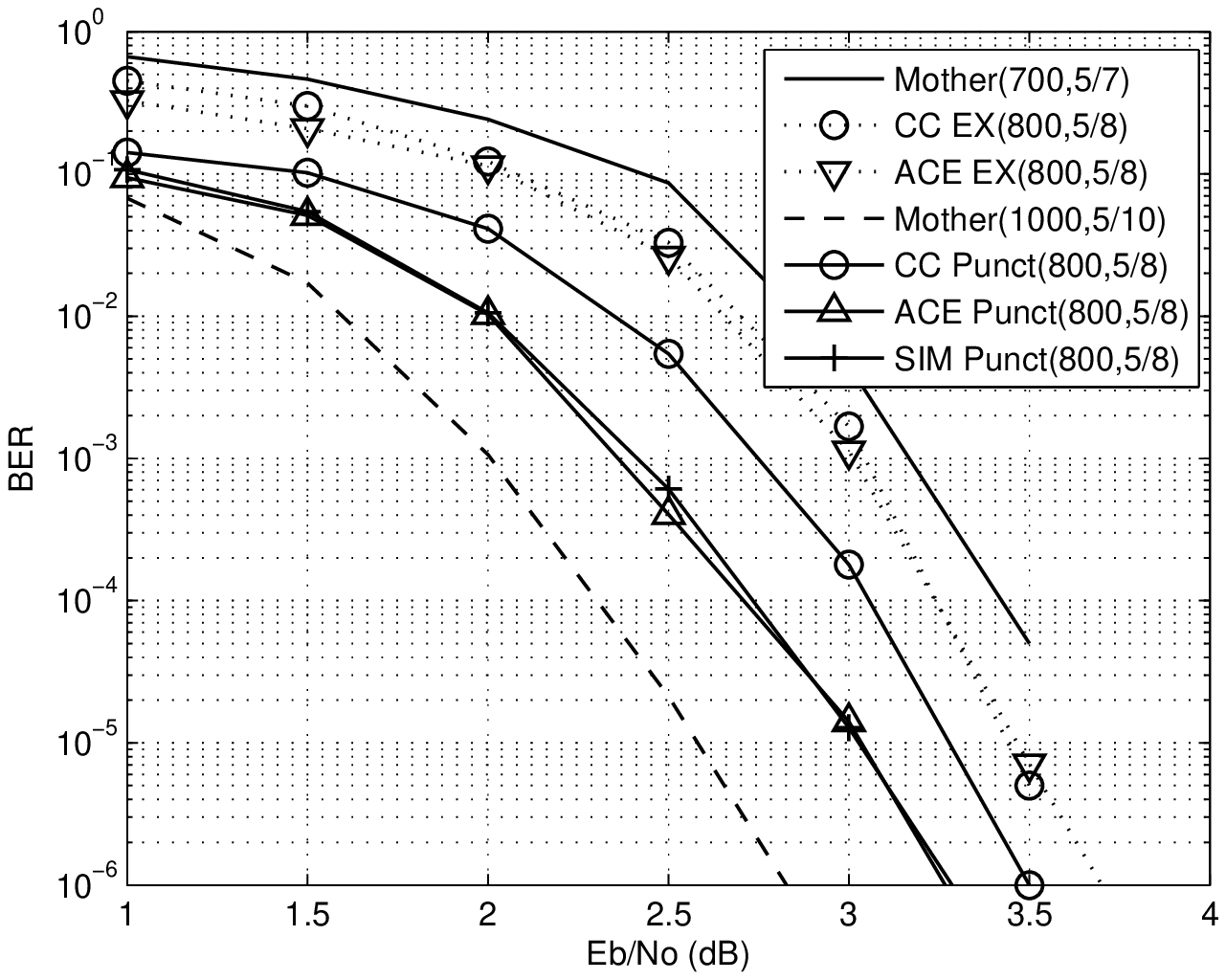} \vspace{-0.5em}\caption{Comparison of the
proposed extension schemes with the proposed puncturing schemes at a
high rate, $5/8$, for irregular LDPC code. }\label{fig:RClhigh}
\end{center}
\end{figure}

\begin{figure}[!htb]
\begin{center}
\def\epsfsize#1#2{0.95\columnwidth}
\epsfbox{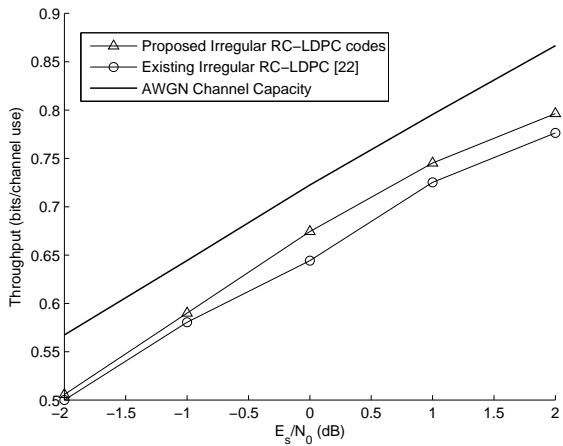} \vspace{-0.5em}\caption{Comparison of the
proposed irregular RC-LDPC codes with the irregular RC-LDPC code
\cite{Yazdani} in system throughput. The capacity of AWGN channel is
also included.}\label{fig:throughput}
\end{center}
\end{figure}

Fig. \ref {fig:RClow} shows that the proposed extension
schemes outperform the proposed puncturing
schemes at low rates. Both CC-based and ACE-based extensions
originates from the mother code, C, while all punctured schemes are
from the mother code $(M=500, N=1400, R=5/14)$. On the other hand, Fig.
\ref{fig:RClhigh} shows that at a high rate, $5/8$ of the punctured
codes (mother code C) offer better performance as compared to the
proposed extension codes whose mother code has $M=500, N=700,
R=5/7$. To illustrate the overall performance of
the proposed RC-LDPC codes, we finally compare the proposed RC-LDPC codes with the existing RC-LDPC family in the system throughput \cite
{Mantha} as shown in Fig. \ref{fig:throughput}, in which $E_b$ in
previous figures is replaced by $E_s$, the average energy per
transmitted symbol. Fig. \ref{fig:throughput} shows that the
proposed RC-LDPC codes are superior to existing RC codes
\cite{Yazdani} and can approach channel capacity.

\section{Conclusion}\label{sec:Sum}

In this paper, we have investigated irregular RC-LDPC codes from
both puncturing and extension perspectives. By applying
counting cycle algorithms, the ACE spectrum and exhaustive searches,
three puncturing schemes as well as two extension schemes have been
devised. All proposed schemes manage to achieve various resulting
rates, and at the same time provide better performance than
existing methods. Simulation results have shown that the proposed
extension designs are suitable for creating RC-LDPC with low rates
$(R<0.5)$ and ACE-based extension performs better than the CC-based
extension. On the other hand, the puncturing designs are preferred
for codes with high rates. With the additional improvement, the ACE
puncturing has been proven to generate the optimal puncturing
pattern and slightly outperform simulation-based puncturing.
As a consequence, taking advantage of a combined
puncturing/extension strategy, we have devised algorithms to
generate RC-LDPC codes with a wide range of rates $(0.1< R <0.9)$.


\begin{thebibliography}{100}


\bibitem{Hagenauer}
J. Hagenauer, "Rate-compatible punctured convolutional codes (RCPC
codes) and their applications," \textit {IEEE Trans. Commun.}, vol.
36, no. 4, pp. 389-400, Apr. 1988.

\bibitem{Yu}
S. Lin and P. S. Yu, "A hybrid ARQ scheme with parity retransmission
for error control of satellite channels," \textit {IEEE Trans.
Commun.}, vol. 30, no. 7, pp. 1701-1719, Jul. 1982.

\bibitem{Rowitch}
D.N. Rowitch, L.B. Milstein, "On the performance of hybrid FEC/ARQ
systems using rate compatible punctured turbo (RCPT) codes," ,
\textit {IEEE Transactions on Communications}, vol.48, no.6,
pp.948-959, Jun 2000.

\bibitem{Gallager}
R. G. Gallager, "Low-density parity check codes," \textit {IRE
Trans. Inf. Theory.}, vol. 39, no. 1, pp. 37-45, Jan. 1962.

\bibitem{Mackay}
D. J. C. Mackay and R. M. Neal, "Near Shannon limit performance of
low density parity check codes," \textit {Electron. Lett.}, vol. 33,
no. 6, pp. 457-458, Mar. 1997.

%\bibitem{Shokrollahi}
%A. Shokrollahi and R. Storn, "Design of efficient erasure codes with
%different evolution," in \textit {Proc. 2000 IEEE International
%Symposium on Information Theory (ISIT)}, pp. 5, Jun 2000.

\bibitem{LDPCcapacity}
T. J. Richardson, M. A. Shokrollahi, R. L. Urbanke, "Design of
capacity-approaching irregular low-density parity-check codes,"
\textit {IEEE Trans. Inf. theory.}, vol. 47, no. 2, pp. 619 - 637,
Feb. 2001.

\bibitem{Forney1}
Sae-Young Chung, G.D. Jr.Forney, T.J. Richardson, R. Urbanke, "On
the design of low-density parity-check codes within 0.0045 dB of the
Shannon limit," \textit {IEEE Communications Letters}, vol.5, no.2,
pp.58-60, Feb 2001.

\bibitem{Proietti}
C. Di, D. Proietti, I. E. Telatar, T. Richardson, and R. Urbanke,
"Finite-length analysis of low-density parity-check codes on the
binary erasure channel," \textit {IEEE Trans. Infor. Theory}, vol.
48, pp. 1570-1579, Jun. 2002.

\bibitem{Ryne1}
M. Yang, W.E. Ryan, Li Yan, "Design of efficiently encodable
moderate-length high-rate irregular LDPC codes," \textit {IEEE
Transactions on Communications}, vol.52, no.4, pp. 564- 571, April
2004.
\bibitem{dopeg_vtc}
C. T. Healy and R. C. de Lamare, ``Decoder optimised progressive
edge growth algorithm", \textit{Proc. IEEE 73rd Vehicular Technology
Conference (VTC Spring)}, 2011.

\bibitem{qc-dopeg}
C. T. Healy and R. C. de Lamare, ``Quasi-cyclic low-density
parity-check codes based on decoder optimised progressive edge
growth for short blocks", \textit{Proc. IEEE International
Conference Acoustics, Speech and Signal Processing (ICASSP)}, 2012,
pp. 2989-2992.

\bibitem{dopeg_cl} C. T. Healy and R. C. de Lamare,
``Decoder-optimised progressive edge growth algorithms for the
design of LDPC codes with low error floors",  \textit{IEEE
Communications Letters}, vol. 16, no. 6, June 2012, pp. 889-892.

\bibitem{peg_bf_iswcs}
A. G. D. Uchoa, C. T. Healy, R. C. de Lamare, R. D. Souza, ``LDPC
codes based on progressive edge growth techniques for block fading
channels",  \textit{Proc. 8th International Symposium on Wireless
Communication Systems (ISWCS)}, 2011, pp. 392-396.

\bibitem{gqcpeg}
A. G. D. Uchoa, C. T. Healy, R. C. de Lamare, R. D. Souza,
``Generalised Quasi-Cyclic LDPC codes based on progressive edge
growth techniques for block fading channels",  \textit{Proc.
International Symposium Wireless Communication Systems (ISWCS)},
2012, pp. 974-978.

\bibitem{peg_bf_cl}
A. G. D. Uchoa, C. T. Healy, R. C. de Lamare, R. D. Souza, ``Design
of LDPC Codes Based on Progressive Edge Growth Techniques for Block
Fading Channels", \textit{IEEE Communications Letters}, vol. 15, no.
11, November 2011, pp. 1221-1223.

%\bibitem{Yue}
%G. Yue, B. Lu and X. Wang, "Analysis and design of finite-length
%LDPC codes," \textit {IEEE Trans. Vehicular Technology.}, vol. 56,
%no. 3, May 2007.

\bibitem{Li}
J. Li and K. R. Narayanan, "Rate-Compatible Low Density Parity Check
(RC-LDPC) Codes for Capacity-Approaching ARQ Schemes in Packet Data
Communications," \textit {Proceeding of International Conference on
on Communications, Internet and Information Technology (CIIT)}, US
Virgin Islands, pp. 201-206, Nov. 2002.

\bibitem{Ha}
Jeongseok Ha, Jaehong Kim, S.W. McLaughlin, "Rate-compatible
puncturing of low-density parity-check codes," \textit {IEEE Trans.
Infor. Theory}, vol.50, no.11, pp. 2824- 2836, Nov. 2004.

\bibitem{Kim}
J. Ha, J. Kim, D. Klinc, and S. W. Mclaughlin, "Rate-compatible
punctured low-density parity-check codes with short bolck length,"
\textit {IEEE Trans. Infor. Theory}, vol. 52, no. 2, pp. 728-738,
Feb. 2006.

\bibitem{Park}
H. Y. Park, J. W. Kang, K. S. Kim, and K. C. Whang, ``Efficient
puncturing method for rate-compatible low-density parity-check
codes," \textit {IEEE TWC}, pp. 3914-3919, Nov. 2007.

\bibitem{Mclaughlin}
Jaehong Kim, A. Ramamoorthy, S. Mclaughlin, "The design of
efficiently-encodable rate-compatible LDPC codes," \textit {IEEE
Transaction on Communications}, vol.57, no.2, pp.365-375, February
2009.

\bibitem{Bhushan}
M. El-khamy, J. Hou, and N. Bhushan, ``Design of rate-compatible
structured LDPC codes for hybrid ARQ applications," \textit {IEEE
JSAC}, pp. 965-973, Aug. 2009.

%\bibitem{Saeedi}
%H. Saeedi, H. Pishro-Nik, and A. H. Banihashemi, ``Successive
%maximization for the systematic design of universally capacity
%approaching rate-compatible sequences of LDPC code ensembles over
%binary-input output-symmetric memoriless channels," \textit {IEEE
%TCOM}, pp. 1807 - 1819, July 2011.

\bibitem{Pishro-Nik}
H. Pishro-Nik and F. Fekri, "Result on punctured low-density
parity-check codes and improved iterative decoding techniques," in
\textit {Proc. 2004 IEEE Infor. Theory Workshop}, pp. 24-29, San
Antonio, TX, Oct. 2004.

%\bibitem{Fekri}
%Hossein Pishro-Nik, Faramarz Fekri, ``Results on Punctured
%Low-Density Parity-Check Codes and Improved Iterative Decoding
%Techniques," \textit {IEEE Transaction on Information Theory},
%vol.53, no.2, pp.599-614, Feb. 2007.

\bibitem{Vellambi}
B. N. Vellambi and F. Fekri, ``Finite-length rate-compatible LDPC
codes: a novel puncturing scheme," \textit {IEEE Transaction on
Communications}, vol. 57, no. 2, pp. 297-301, Feb. 2009.

\bibitem{Yazdani}
M. R. Yazdani and A. H. Banihashemi, ``On comstruction of
rate-compatible low-density parity-check codes," \textit {IEEE
Commun. Lett.}, vol. 8, no. 3, pp. 159-161, Mar. 2004.

%\bibitem{Tian}
%T. Tian, C. Jones, ``Construction of rate-compatible LDPC codes
%utilizing information shortening and parity puncturing," \textit
%{EURASIP Journal on Wireless Communications and Networking.}, Vol.
%2005, no. 5, Oct. 2005.

\bibitem{Liu}
Jingjing Liu, R.C. de Lamare, ``Novel intentional puncturing schemes
for finite-length irregular LDPC codes," \textit {17th International
Conference on Digital Signal Processing (DSP)}, vol., no., pp.1-6,
6-8 July 2011.

\bibitem{Halford}
T. R. Halford, K. M. Chugg, "An algorithm for counting short cycles
in bipartite graph," \textit {IEEE Trans. on Infor. Theory}, vol.
52, no. 1, pp. 287-292, Jan. 2006.

\bibitem{Tao}
Tian Tao, C.R. Jones, J.D. Villasenor, R.D. Wesel, ``Selective
avoidance of cycles in irregular LDPC code construction," \textit
{IEEE Transaction on Communications},, vol.52, no.8, pp. 1242- 1247,
Aug. 2004.

\bibitem{Vukobratovic}
D. Vukobravic, V. Senk, "Evaluation and design of irregular LDPC
codes using ACE spectrum," \textit {IEEE Trans. Commun.}, vol. 57,
no. 8, pp. 2272 - 2279, Aug. 2009.

\bibitem{Liu2}
Jingjing Liu, R.C. de Lamare, ``Finite-length rate-compatible LDPC
codes based on extension techniques," \textit {8th International
Symposium on Wireless Communication Systems (ISWCS)}, vol., no.,
pp.41-45, 6-9 Nov. 2011.

\bibitem{Karimi}
M. Karimi, A.H. Banihashemi, ``A message-passing algorithm for
counting short cycles in a graph," \textit {IEEE Information Theory
Workshop (ITW)}, vol., no., pp.1-5, 6-8 Jan. 2010.

\bibitem{Xiao}
Hua Xiao, A.H. Banihashemi, ``Improved progressive-edge-growth (PEG)
construction of irregular LDPC codes," \textit {IEEE Communications
Letters}, vol.8, no.12, pp. 715- 717, Dec. 2004.

\bibitem{Hu}
Y. Hu, E. Eleftheriou and D. M. Arnold, "Regular and irregular
progressive edge-growth tanner graph," \textit {IEEE Transaction on
Information Theory}, vol. 51, no. 1, pp. 386-398, Jan. 2005.

\bibitem{Tom}
T. J. Richardson, ``Error Floor of LDPC Codes," in \textit {Proc.
41st Annu. Allerton Conf. on Communication, Control, and Computing},
Urbana-Champaign, Oct. 2003, pp. 1426-1435.

\bibitem{Mantha}
 R. Mantha, F.R. Kschischang, ``A capacity-approaching hybrid ARQ
scheme using turbo codes," \textit {Global Telecommunications
Conference, 1999. GLOBECOM '99,} vol.5, no., pp. 2341- 2345 vol.5,
1999.

\bibitem{ra-ldpc}
A. G. D. Uchoa, C. T. Healy, R. C. de Lamare , ``Repeat Accumulate
Based Constructions for LDPC Codes on Fading Channels",
\textit{Proc. International Symposium on Wireless Communication
Systems (ISWCS 2013)}, 2013.

\bibitem{vfap_cl} J. Li and R. C. de
Lamare, ``Low-Latency Reweighted Belief Propagation Decoding for
LDPC Codes",  \textit{IEEE Communications Letters}, vol. 16, no. 10,
October 2012,  pp. 1660 - 1663.

\bibitem{spa}
R.C. de Lamare, R. Sampaio-Neto, "Minimum mean-squared error
iterative successive parallel arbitrated decision feedback detectors
for DS-CDMA systems", IEEE Trans. Commun., vol. 56, no. 5, May 2008,
pp. 778-789.

\bibitem{stspadf}
Y. Cai and R. C. de Lamare, "Adaptive Space-Time Decision Feedback
Detectors with Multiple Feedback Cancellation", \textit{IEEE
Transactions on Vehicular Technology}, vol. 58, no. 8,  October
2009, pp. 4129 - 4140.


\bibitem{mdfpic}
P. Li and R. C. de Lamare, "Adaptive Decision-Feedback Detection
With Constellation Constraints for MIMO Systems", \textit{IEEE
Transactions on Vehicular Technology}, vol. 61, no. 2, 853-859,
2012.

\bibitem{mbdf} R. C. de Lamare, "Adaptive and Iterative
Multi-Branch MMSE Decision Feedback Detection Algorithms for
Multi-Antenna Systems", \textit{IEEE Transactions on Wireless
Communications}, vol. 14, no. 2, February 2013.




\end{thebibliography}
\end{document}